\documentclass{emulateapj}
\usepackage{amsmath}

\long\def\symbolfootnote[#1]#2{\begingroup%
\def\thefootnote{\fnsymbol{footnote}}\footnote[#1]{#2}\endgroup}

\newcommand{\beq}{\begin{equation}}
\newcommand{\eeq}{\end{equation}}
\newcommand{\bea}{\begin{eqnarray}}
\newcommand{\eea}{\end{eqnarray}}

\newcommand{\vsmax}{\hat v_{s,{\rm max}}}
\newcommand{\vvec}{{\mathbf v}}
\newcommand{\ddalpha} { {\partial \over \partial \alpha} } 
\newcommand{\vphi}{{v_\varphi}}
\newcommand{\Lphi}{{\ell}_\varphi}
\newcommand{\Gammaphi}{{\Gamma_\varphi}}

\slugcomment{ApJ, in press}

\shorttitle{Oblique Supernova Shock Breakout}
\shortauthors{Matzner, Levin, \& Ro}

\begin{document}

\title{Oblique Shock Breakout in Supernovae and Gamma-Ray Bursts: I.  Dynamics and Observational Implications}

\author{Christopher D. Matzner\altaffilmark{1}, Yuri Levin\altaffilmark{2}, \& Stephen Ro\altaffilmark{1}}
\altaffiltext{1}{Department of Astronomy \& Astrophysics, University of Toronto, 50 St. George St., Toronto, ON M5S 3H4, Canada}
\altaffiltext{2}{Monash Centre for Astrophysics, Monash University, Clayton, VIC 3800, Australia}
\email{matzner@astro.utoronto.ca}
\begin{abstract}
In a non-spherical stellar explosion, non-radial motions become important near the stellar surface.  For realistic deviations from spherical symmetry, non-radial flow dramatically alters the dynamics and emission of shock emergence on a significant fraction of the surface.  The breakout flash is stifled, ejecta speeds are limited, and matter is cast sideways.  Non-radial ejection allows for collisions outside the star, which may engender a new type of transient.  Strongly oblique breakouts are most easily produced in compact stellar progenitors, such as white dwarfs and stripped-envelope core collapse supernovae.  
 We study the shock structure and post-shock acceleration using conservation laws, a similarity analysis, and an approximate theory for oblique shocks.   The shock is likely to extend vertically from the stellar surface, then kink before joining a deep asymptotic solution.   Outflow from the region crossed by an oblique shock is probably unsteady and may affect the surface ahead of the main shock.  We comment on the implications for several notable explosions in which the non-spherical dynamics described in this paper are likely to play an important role. We also briefly consider relativistic and superluminal pattern speeds. 
\end{abstract}

%\date{}


\maketitle

%%%%%%%%%%%%%%%%%%%%%%%%%%%%%%%%%%%%%%%%%%%%%%%%%%%%%%%%%%%%%%%%%%%%
\section{Introduction} \label{S:intro}
The arrival of a supernova explosion at the surface of its host star sparks several transient and high-energy phenomena.   It marks with a flash the beginning of the electromagnetic display, as the post-shock radiation field is revealed to observers \citep{1978ApJ...223L.109K,1989ASPRv...8....1I}.  Thanks to a whip-like acceleration in progressively more diffuse layers of the stellar envelope \citep{1956SPhD....1..223G}, shock emergence also involves the fastest ejecta -- those which dominate the early light curve and early radio or X-ray emission in circumstellar interactions \citep{1988A&A...192..221L,2006A&A...449..171N}.  Sufficiently compact and energetic supernovae generate relativistic ejecta \citep{1999ApJ...510..379M,2001ApJ...551..946T}, which can create $\gamma$-ray or X-ray flashes, accelerate high-energy cosmic rays \citep{2008ApJ...673..928B}, and spall light nuclei \citep{2002ApJ...581..389F,2006ApJ...643L.115N} in a collision with matter around the star.   Breakouts or breakout-induced interactions have been invoked for the transients observed in supernovae 1998bw \citep{1999ApJ...510..379M,2001ApJ...551..946T} and 2008D \citep{2008ApJ...683L.135C,2008MNRAS.388..603L}, for the ionization of a circumstellar nebula in 1987A \citep{1988A&A...192..221L}, and for the excitation of thermal dust echoes in the Cas A remnant \citep{2008ApJ...685..976D}.

Indeed, relativistic shock breakouts \citep{2012ApJ...747...88N}, relativistic ejecta-wind interactions \citep{2001ApJ...551..946T,1999ApJ...510..379M}, or shock breakouts through thick winds  \citep{2011MNRAS.414.1715B,2012ApJ...759..108S} have been invoked to explain the low-luminsoity gamma-ray bursts and smooth, single-peak $\gamma$-ray bursts, as well as X-ray flashes which appear to constitute a common, weakly beamed population distinct from the cosmological $\gamma$-ray burst population \citep{1998ApJ...506L.105B,2006ApJ...645L.113C,2006Natur.442.1014S,2007A&A...465....1D,2007ApJ...657L..73G,2007ApJ...661..394L,2007ApJ...662.1111L,2007MNRAS.382L..21C,2008A&A...484..143F,2009MNRAS.392...91V}.

The dynamics involved in shock acceleration are well understood in both the non-relativistic \citep{sakurai60,1999ApJ...510..379M} and relativistic \citep{2006ApJ...643..416P,2005ApJ...627..310N} limits and accurate approximations exist for the shock velocity and post-shock acceleration \citep{1999ApJ...510..379M} and transition from Newtonian to relativistic flow \citep{2001ApJ...551..946T}.  Great progress has also been made in understanding the radiation dynamics which terminate shock acceleration and release the breakout flash \citep{2000ApJ...532.1132B,2010ApJ...716..781K, 2011ApJ...742...36S,2012ApJ...747..147K,2012ApJ...747...88N,2013arXiv1304.6428S}.  

However, all of these theoretical advances assume strict spherical or planar symmetry.   Supernovae are unlikely to ever be purely spherical, either because the progenitor star is rapidly rotating or tidally distorted \citep[e.g.,][]{1989ApJ...341..867C}, or because asymmetries are necessary or inherent to the explosion mechanism \citep{2001A&A...368..311F,2003ApJ...584..971B,2009ApJ...703.1464F} -- jet-driven explosions \citep[e.g.,][]{1999ApJ...524L.107K,2003MNRAS.345..575M} being an extreme example.  Moreover, deviations from spherical symmetry grow wherever the shock accelerates within the stellar envelope (while also damping in zones of deceleration).  

One aspect of the spherical theory strongly suggests that shock breakout should be strengthened when spherical symmetry is broken.  A key feature of spherical explosions is the very steep decline of kinetic energy with ejecta velocity: for instance, the kinetic energy above some high final velocity $v_f$ scales as $v_f^{-5.2}$ for sufficiently high (but non-relativistic) $v_f$.   Small changes in the shock strength therefore lead to much larger variations in the high-velocity ejecta.  Extrapolating this rule, one might expect angular variations in an explosion to be accentuated by shock breakout, leading to more vigorous transients in some directions and even enhancing the high-velocity share of the total energy budget \citep{2001ApJ...551..946T,2003MNRAS.345..575M}.  

But this argument still neglects non-radial motions, which arise naturally in non-spherical explosions.  Our goal in this paper is to account for such motions and how they alter the dynamics of shock emergence.  In fact, the differences from radial flow can be dramatic.  As we shall see,  non-radial flow imposes a new upper limit on the ejecta speed, which is twice the local pattern speed.  Moreover, where this new limit matters, the flow is deflected towards the stellar surface and the photon flash which accompanies breakout is suppressed.

We classify asymmetrical shock breakouts (\S~\ref{S:ObliqueSBO_NonrelVsRel}), briefly considering relativistic and superluminal pattern speeds.  We present analytical results for the case of non-relativistic pattern speeds (\S~\ref{S:NR-ShockStructure} and \S~\ref{S:NR-SimilarityAnalysis}), and consider the observational implications (\S~\ref{S:implications}), and specific supernovae in \S~\ref{S:discussion}.  Numerical results for the non-relativistic case follow in a second paper (Salbi et al.\ 2013, hereafter Paper 2).  

\section{Oblique shock breakout: asymptotic non-relativistic and superluminal flow } \label{S:ObliqueSBO_NonrelVsRel} 

We begin our study of aspherical explosions by reviewing the spherical theory.  Within a spherical explosion, the dynamics of shock motion enter a regime of planar motion near the stellar surface.  If the stellar density distribution is polytropic, $\rho_0(y<0) = \rho_{h} (-y/R_*)^n$, where $y = r-R_*$ is the altitude relative to the stellar surface, then the second-type similarity solution of  \cite{sakurai60} applies, in which 
\beq\label{vs-S60} 
\hat v_s(y<0) = v_1 (-y/R_*)^{-\lambda}. 
\eeq 
Variables with hats refer to a spherical or planar flow. 

This shock acceleration law is equivalent to the density-velocity scaling $\hat v_s\propto \rho_0^{-\beta}$ if $\lambda = n \beta$.    After the shock passes through, each fluid element accelerates to a terminal speed $v_f(m)$ which, in the planar solution, is about twice the shock velocity $v_s(m)$ it experienced.  In a spherical star, elements from deeper layers suffer spherical effects which rapidly limit the post-shock acceleration factor $\hat v_f(m)/\hat v_s(m)$. 

In spherical theory, the shock-strength coefficient $v_1$ reflects the scale $v_*$ set by energy conservation, but also the tendency for shocks to accelerate where the density declines rapidly.  \citet{1999ApJ...510..379M} demonstrate that, if $m(r)$ is the amount of to-be-ejected mass initially enclosed within $r$, the approximation 
\beq \label{vs-MM99} 
\hat{v}_s = C_1 \left(M_{\rm ej}\over m\right)^{1/2}  \left(m\over \rho_0 r^3\right)^\beta v_* 
\eeq 
is accurate to a few percent -- 
even when the coefficient $C_1= 0.794$ is matched to a single spherical blastwave (for $\rho_0\propto r^{-17/7}$, which is characteristic of various progenitors, and adiabatic index $\gamma=4/3$)  and the exponent $\beta=0.19$ is matched to \citet{sakurai60}'s planar solutions for $\gamma=4/3$ (intermediate between  $\beta=0.1858$ [$n=3$], and $\beta=0.1909$ [$n=3/2$]).  Accordingly \[v_1 = C_1 \left(M_{\rm ej}\over \rho_h R_*^3\right)^{\beta} v_*\] to good accuracy in a spherical explosion.  Further accuracy can be attained by adjusting $C_1$  \citep{2001ApJ...551..946T} or $\beta$ \citep{RoMatzner2013} using information about the stellar density distribution. 

Another fundamental feature of spherical explosions is the breakdown of adiabatic theory where the shock optical depth $\tau_s = c/(3\hat v_s)$ exceeds the vertical optical depth of the stellar envelope, $\tau(y) = \kappa y \rho(y)/(n+1)$; this  condition sets the maximum velocity of planar shock breakout, 
\beq\label{vsmax}
\vsmax = v_1 \left[ {3v_1\kappa \rho_h R_*\over (n+1)c}\right]^{\lambda\over 1+n-\lambda},
\eeq
and determines the character of the breakout flash.   For sufficiently fast shocks in sufficiently dense envelopes, those with 
\beq
(2v_1/c)^{\gamma_p/\beta} \kappa \rho_{h} R_* > 2(1+n)/3
\eeq 
(where $\kappa$ is the electron-scattering opacity and $\gamma_p=1+1/n$ the polytropic index), the shock becomes relativistic before breakout, in the sense that $\hat v_s>c/2$ according to equation (\ref{vsmax}). A characteristic distribution of relativistic ejecta is then produced, and a trans-relativistic treatment \citep{2001ApJ...551..946T} is required. 

The essential property of non-spherical explosions is that the strength and timing of the shock  are not constant.  In other words, both the shock strength $v_1$ and the time  of shock emergence $t_{\rm se}$ vary across the stellar surface, and the two-dimensional gradient of $t_{\rm se}$ defines the lateral pattern or phase speed,  $\vphi = |\nabla_2 t_{\rm se}|^{-1}$ at which the emerging shock marches across the surface.   
We refer to this as {\em obliquity}, as it causes the shock normal to become non-radial. 
While the implications of non-simultaneity for the breakout light curve have been considered by  \citet{2010ApJ...717L.154S} and \citet{2011ApJ...727..104C} \citep[and in a preliminary way by][]{2004MNRAS.351..694C},  implications of obliquity for breakout dynamics have not.   

As $\vphi$ sets the velocity scale for an oblique breakout, comparing $\vphi$ to the explosion velocity scales $v_*$ and $\vsmax$ reveals when and where obliquity will be important.   As we shall see in \S 3, obliquity affects the flow when $\hat v_s\simeq \vphi$ and alters it completely in those parts of the atmosphere where the spherically symmetric theory would predict $\hat v_s>\vphi$.   Globally asymmetric flows have $\vphi \lesssim v_*$, so that spherical symmetry is never appropriate; however where $\vphi \gg v_*$ the region affected by obliquity is limited to in a thin outer layer of shock acceleration.   However, when $\vphi > \vsmax$ radiation transfer affects the flow before obliquity can, so that shock dynamics and the radiation flash are little altered.     Being a pattern speed, $\vphi$ can exceed the speed of light, and the spherically symmetric case is realized in the limit $\vphi\rightarrow\infty$.   We provide some examples in \S \ref{S:Examples}, where we estimate $\vphi$ for asymmetric explosions of some standard progenitor models.

Globally-asymmetric flows require global numerical simulations, but we focus our attention on the case $\vphi\gg v_*$ in which 
 a local treatment is possible.  We neglect radiation transport and consider only adiabatic flow; for non-relativistic pattern speeds this restricts us to $\vphi<\vsmax$.  In this sense our analysis refers to an asymptotic limit of oblique breakouts, valid anywhere that shock acceleration is described by \citet{sakurai60}'s planar solution.  
Given this restriction, we divide the problem into non-relativistic ($v_*<\vphi<c$), trans-relativistic ($\vphi\lesssim c$), and superluminal ($\vphi>c$) cases, bearing in mind that a single explosion might encompass all three regimes. 

 In the non-relativistic case, formula (\ref{vs-S60}) implies $\hat v_s(y) = \vphi$ at $y=-\Lphi$ where 
 \beq\label{Lphi} 
 \Lphi = (v_1/\vphi)^{1/\lambda} R_*, 
 \eeq
the size of the region where non-radial motion is imprinted on the breakout flow and the distribution of ejecta.  So long as $\Lphi |\nabla_2\ln v_1| \ll 1$ and $\Lphi |\nabla_2\ln \vphi| \ll 1$, matter meets the shock in steady flow.   We analyze the properties of the non-relativistic post-shock flow in \S~\ref{S:NR-universal-solns} and Paper 2, using $\Lphi$ as the fundamental length scale. 

For superluminal pattern speeds it is no longer useful to consider the comoving frame of motion.  However, in this case 
the shock breakout events in each patch of the stellar surface are spacelike-separated and become simultaneous in a frame moving tangentially along the stellar at speed $v_{\rm boost}=c^2/\vphi$ in the same direction as $\nabla_2 t_{\rm se}$.    When $\vphi \gg c$ so that $v_{\rm boost}\ll c$ this amounts to a simple Galilean transformation of the known planar solution, but when $\vphi \gtrsim c$ the planar solution must be re-computed for an atmosphere initially traveling laterally at speed $v_{\rm boost}$ before it can be transformed back into the star's rest frame.   We leave this calculation to a future paper.

\section{Two-dimensional model and an asymptotic solution for non-relativistic pattern speeds} \label{S:NR-universal-solns}

We wish to characterize in this section that part of the flow which extends outward from the point (or line) of shock breakout, through the zone in which obliquity strongly modifies the shock motion, and into the region where planar symmetry is a good approximation.  We specialize to the combined limit that $\vphi\ll c$ (non-relativistic flow), $\vphi \ll \vsmax$ (adiabatic flow), $\Lphi \ll R_*$ (initially planar flow), and both $\Lphi \ll  |\nabla_2\ln(v_1)|^{-1}$ and $\Lphi \ll  |\nabla_2\ln(\vphi)|^{-1}$ (steady flow).   As remarked above, this combination represents an asymptotic limit of the more general case of non-relativistic pattern speeds, whose solution is a useful guide even in cases where some of these assumptions are broken.   

Moreover, these limits greatly simplify our analysis because the flow is now described by only one scale of length ($\Lphi$), of velocity ($\vphi$), of time ($t_\varphi = \Lphi/\vphi$), and of density ($\rho_\varphi = \rho_0(y=-\Lphi)$).  In our asymptotic problem, the stellar density is initially planar in symmetry and a power-law of depth.  Any curvature and unsteadiness of the shock front involve scales much larger than $\Lphi$, so the shock intersects the stellar surface along a line, and we consider the two-dimensional flow perpendicular to this line (or `point') of breakout. 

For any combination of the adiabatic and polytropic indices ($\gamma$ and $\gamma_p$) there exists a unique, universal solution to the non-relativistic oblique breakout problem to accompany the self-similar (scale-free) planar flow described by \citet{1956SPhD....1..223G}, \citet{sakurai60}, \citet{1999ApJ...510..379M}, and \citet{RoMatzner2013}.   Indeed,  for matter originating well below the obliquity scale ($R_*\gg |y_0(m)|\gg\Lphi$), in which the shock normal is almost radial, this planar solution holds and obliquity is an increasingly minor perturbation.  Figure \ref{fig:NR_breakout} depicts the flow on scales of order $\Lphi$. 

The flow is stationary in a frame of reference moving along the stellar surface at velocity $\vphi$.  We define the $x$ coordinate axis to be orthogonal to the surface normal and the line where the shock breaks the surface. We set $x=0$ along this line, and define $x<0$ upstream and $x>0$ downstream of it.  Because the initial density distribution is effectively planar, the unbroken region of the stellar surface is defined by $y=0, x<0$.  

Steady motion has two immediate implications for the properties of the shock and the post-shock flow.  First, in this frame unshocked matter flows along the stellar surface at speed $\vphi$ before crossing the shock front; therefore $\vphi$ is the maximum velocity of flow normal to the shock front. 

Second, in steady flow, mass and energy travel along streamlines.  Ignoring the initial hydrostatic pressure of the envelope and changes in the gravitational potential on the scale of interest, the Bernoulli function ${\cal B}$, or the ratio of energy flux to mass flux, is constant everywhere in the flow and equal to its value in the inflow: 
\begin{equation} \label{Bernoulli} 
{\cal B} = \frac12 v^2+ {\gamma\over \gamma-1} {P\over \rho}= \frac12 \vphi^2. 
\end{equation} 
Because the pressure $P$ cannot be negative, the pattern speed $\vphi$ is an upper limit to the flow speed in the steady-state frame, and as we expect $P\rightarrow 0$ in the outflow, the terminal outflow velocity in this frame is uniquely $\vphi$.   In the star's rest frame this translates to a maximum ejecta speed of $2 \vphi$, which is achieved if matter is thrown along the stellar surface (in the same direction as the emerging shock).    

These points highlight one of obliquity's major effects: it ends the acceleration of the shock front and post-shock flow in the outer reaches of the star, a process which would otherwise continue until radiative transfer or relativistic effects set in.  This has dramatic implications for the production of a photon flash and all the other energetic phenomena surrounding shock emergence, as we discuss in \S~\ref{S:discussion}.    (There is an analogous result for trans-relativistic flows; see \S~\ref{S:Transrel}.) 

A couple additional results follow from the fact that the shock and the post-shock flow obey the known planar solution in regions unaffected by obliquity, curvature, or radiation diffusion.  Flow well below obliquity scale is planar and directed vertically in the star's rest frame; its shock motion is described by equation (\ref{vs-S60}) and related to $\Lphi$ by (\ref{Lphi}).  Mapping this motion into the steady-state frame, the surface of the shock $(x_s,y_s)$ is given by 
\beq \label{Shock_Shape_in_Deep_Zone_of_Steady_Flow}
\left(-y_s\over \Lphi\right)^{\lambda+1} \rightarrow (\lambda+1){x_s - x_{s0}\over \Lphi} ~~~{\rm for}~ y_s\ll-\Lphi.
\eeq 
Here $x_{s0}$ is an offset of order $\Lphi$ which accommodates the fact that obliquity affects the flow in the region $y\sim-\Lphi$, so that breakout ($x=y=0$) does not occur exactly where one would predict by extrapolating from planar flow. 

Now, consider the end state of this deep matter -- that which originates below the obliquity scale, yet in a region where the initial density is still effectively planar ($R_*\gg |y_0|\gg \Lphi$).  The terminal vertical velocity in planar flow is $\hat v_f(m) = C_2 \hat v_s(m)$ where $C_2\simeq 2$ (to be precise $C_2= [2.17,2.03,1.94, 1.85]$ when $\gamma=4/3$ and $n=[1.5, 3, 7, \infty]$; see \citealt{RoMatzner2013}), and our definitions imply $\hat v_s(y_0) = \vphi (-\Lphi/y_0)^\lambda$.  Because the flow is planar below the obliquity scale, $v_s(y_0)=\hat v_s(y_0)$ for $|y_0|\gg -\Lphi$.   The vertical component of this terminal flow is the same when viewed in the steady-state frame, but from equation (\ref{Bernoulli}) we know that the terminal speed is $\vphi$ in that frame.   For these statements to be consistent, 
\beq\label{Terminal_vx_steady_frame}
v_{xf}' = \left[1-C_2^2 \left(\Lphi\over -y_0\right)^{2\lambda}\right]^{1/2} \vphi; 
\eeq
prime denotes the steady-state frame, and the positive of the two possible solutions is correct.  In the star's rest frame,  
\beq\label{Terminal_vx_star's_frame}
v_{xf} = -\left\{1-\left[1-C_2^2 \left(\Lphi\over -y_0\right)^{2\lambda}\right]^{1/2} \right\} \vphi. 
\eeq
This is valid only in the limit $|y_0|\gg\Lphi$, and breaks down entirely for $|y_0|< C_2^{1/\lambda} \Lphi$.    Where it is valid it implies that, in the star's rest frame, ejecta are deflected by an angle 
\begin{equation} \label{deflection_angle}
\arctan\left(\hat v_s\over \vphi-\sqrt{\vphi^2-\hat v_s^2}\right)
\end{equation} 
from the radial direction.  

Recalling our choice for the $x$ coordinate, this means that the deep matter is thrown slightly forward relative to the advancing shock in the frame of the star;  this is logical, considering the tilt of the shock front.  In the steady-state frame, equation (\ref{Terminal_vx_steady_frame}) shows there is a characteristic exit angle which depends upon the depth from which this matter was excavated.   However, these results rely on the vertical flow adhering to the planar solution, so they are not valid for small values of the initial depth $|y_0|/\Lphi$ or when the deflection from vertical becomes significant.   For matter strongly affected by obliquity we must consider flow on the scale of $\Lphi$ or even closer to the surface, which we address in \S~\ref{S:NR-ShockStructure} and \S~\ref{S:NR-SimilarityAnalysis}. 

%%%%%%%%%%%%%%%%%%%%%%%%%%%%%%%%%%%%%
\subsection{Shock structure} \label{S:NR-ShockStructure} 

A full description of the oblique breakout flow on the scale of $\Lphi$ requires two-dimensional numerical simulations like the ones we shall present in Paper 2.  However, we can gain some insight by considering the approximate theory for oblique shocks in inhomogeneous media developed by \citet{1964PThPh..32..207I}. These authors first solve for the flow induced by a shock front as it runs at some angle over a contact discontinuity between two uniform regions.  Then, taking the discontinuity to be infinitesimal and making some assumptions (discussed below) about the downstream flow, they derive the corresponding change of shock angle.  Taking the upstream matter to have a polytropic pressure-density relation and integrating over the shock surface, they predict the following relation between the shock angle ($\mu=-\sin(\alpha_s)$ for local shock angle $\alpha_s$ relative to the stellar surface -- see \S \ref{S:NR-SimilarityAnalysis}) and the upstream density $\rho_0$:
\beq\label{Ishizuka_Shock_Relation}
\begin{split}
{\rho_0\over \rho_\varphi} = &{D\over1-\mu^2} \left( {\sqrt{8\mu^2}- \sqrt{8 \mu^2-1} \over \sqrt{8\mu^2} + \sqrt{8 \mu^2-1}} \right)^{4/\sqrt{7}} \\ 
&\times  
\left( {\sqrt{7\mu^2}+ \sqrt{8 \mu^2-1}\over \sqrt{7\mu^2}-\sqrt{8 \mu^2-1}} \right)^{\sqrt{2}}
\end{split}
\eeq
for an arbitrary constant $D$. 
We have specialized their equation (4.5) to $\gamma=4/3$.   Because $\rho_0/\rho_\varphi=(-y/\Lphi)^n$, equation (\ref{Ishizuka_Shock_Relation}) fixes the shape of the shock front -- its angle as a function of depth, which can be integrated to give $x_s(y_s)$ -- up to the scaling factor $D$.    

A remarkable prediction of this theory is that the shock cannot extend below a minimum angle, $\mu=1/\sqrt{8}$ (or more generally, $\mu^2=(\gamma-1)/(2\gamma)$), and that the shock cannot reach the surface at this angle.  This implies that once the shock has bent within $20.7^\circ$ of the vertical, it must jump discontinuously in its orientation  (emitting two weak shocks into the downstream flow; see
\citealt{1959flme.book.....L}) and readjust to the angle at which it meets the surface.  The similarity analysis of \S \ref{S:NR-SimilarityAnalysis} implies that this terminal angle is most likely to be vertical, i.e., at right angles to the stellar surface.  

\begin{figure}
 \includegraphics[width=9cm, angle=0]{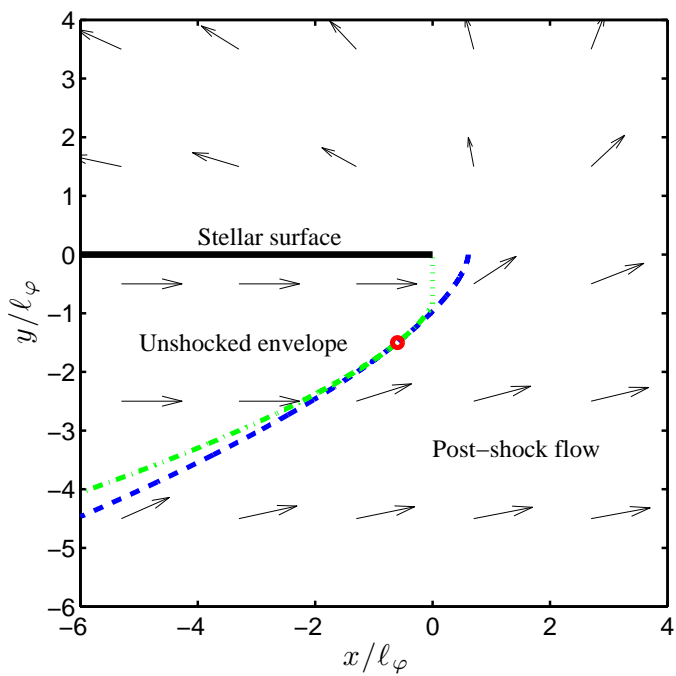}
\caption{Schematic of the oblique breakout flow for non-relativistic pattern speeds in the asymptotic limit $v_*\ll \vphi\ll \vsmax$.  In a steady-state frame which follows the shock along the stellar surface, matter flows from the left across the shock at speed $\vphi$ and is diverted into a spray whose terminal velocity is $\vphi$ in all directions.  Two approximate forms are shown  for the shape of the shock front: an extrapolation of the self-similar shock acceleration law (eq.~(\ref{Shock_Shape_in_Deep_Zone_of_Steady_Flow}) with $x_{0s}=0.61\Lphi$, blue dashed line), which is valid in the limit $y\ll -\Lphi$, and the \citet{1964PThPh..32..207I} approximation (eq.~(\ref{Ishizuka_Shock_Relation}) with $D^n=0.80$, green dash-dot line). These constants are chosen to match values and slopes at $y=-1.5\Lphi$ (red circle).  For both we assume a polytropic structure with $n=3$ and adiabatic index $\gamma=4/3$.   The \citeauthor{1964PThPh..32..207I} theory does not continue past a critical shock angle of $20.7^\circ$ from the vertical; we assume a transition to a perfectly vertical shock at that point (green dotted line), for reasons discussed in the similarity analysis of \S~\ref{S:NR-SimilarityAnalysis}. In the star's rest frame most of the flow is radial, except that matter from a layer of width $\sim \Lphi$ is diverted to the left.  }
\label{fig:NR_breakout}
\end{figure}

Although \citeauthor{1964PThPh..32..207I}'s theory makes definite predictions, it is based on assumptions which cannot be correct in detail.  In particular, it assumes that the post-shock flow is hydrostatic (their eq.\ 3.1b), whereas in fact stellar gravity is negligible in the downstream flow.  Equivalently it ignores the arrival of sound waves from the down-stream regions of the flow, a topic considered by \citet{Moeckel52}.  As a consequence, the shock structure defined by equation (\ref{Ishizuka_Shock_Relation}) does not approach the self-similar planar solution (eq.~\ref{Shock_Shape_in_Deep_Zone_of_Steady_Flow}) for $y\ll -\Lphi$.    
Indeed, the hydrostatic assumption underestimates the gradient of post-shock pressure responsible for accelerating the ejecta.  This leads to a predicted shock angle which is increasingly too oblique at greater depths, as can be seen by comparing equation (\ref{Ishizuka_Shock_Relation}) to the power-law asymptote in Figure  \ref{fig:NR_breakout}. 

Despite this shortcoming, we anticipate that the \citeauthor{1964PThPh..32..207I} model captures features of the shock structure around the obliquity scale, where $\mu$ is relatively constant and the post-shock pressure is approximately a multiple of the pre-shock density: it then resembles in form, if not in magnitude, a hydrostatic distribution.    In Paper 2 we will test the hypothesis that the shock front adheres to equation (\ref{Shock_Shape_in_Deep_Zone_of_Steady_Flow}) deep within the star, but transitions smoothly to the prediction of equation (\ref{Ishizuka_Shock_Relation}) at some reference depth of order $\Lphi$. (A model of this type can be traced by switching from the blue to the green curve at the red circle in Figure \ref{fig:NR_breakout}.)

%%%%%%%%%%%%%%%%%%%%%%%%%%%%%%%%%%%%%
\subsection{Similarity Analysis Around the Breakout Point} \label{S:NR-SimilarityAnalysis} 

So long as radiation is well trapped on the scales where obliquity becomes important ($\vphi\ll \vsmax$), the flow remains adiabatic to much smaller scales ($|y|\ll \Lphi$).  In this case it is reasonable to expect the dynamics to become self-similar as we approach the point (or line) of breakout.  The form of this self-similarity is different from that of planar shock breakout, as it involves an angular variation rather than a temporal one.  
If we suppose that the explosion shock approaches the stellar surface at some definite angle and that the matter upstream of the shock is undisturbed prior to its arrival, then each streamline should sweep through an identical pattern of states as a function of angle.   But, as we shall see, a steady flow of this form is not actually possible.  This result, although puzzling, invites speculation about how the stellar surface is breached. 

We work in polar coordinates $(\varpi,\alpha)$ around the point where the shock meets the surface.   The initial density distribution is $\rho_0 \propto (-y)^n$ for $y=\varpi\sin \alpha <0$, and zero for $y>0$.  Cold matter with this distribution flows to the right at speed $\vphi$, 
and meets a shock of angle $\alpha_s$.  The shock is in the lower half-plane ($-\pi<\alpha_s<0$).  Just before the shock the velocity ${\mathbf v}={\mathbf v}_0  =  (v_{0\varpi},v_{0\alpha}) = (\cos\alpha_s,\sin\alpha_s)\vphi$.   The shock compression factor is $\chi = (\gamma+1)/(\gamma-1)$.  Immediately after the shock ${\mathbf v} = {\mathbf v}_1$ where $v_{1\varpi} = v_{0\varpi} =  \vphi\cos\alpha_s $ and $v_{1\alpha} = v_{0\alpha}/\chi =  \vphi (\sin\alpha_s)/\chi$.  Just behind the shock, the isothermal sound speed $c_i=(P/\rho)^{1/2}$ satisfies $c_{i1}^2 = P_1/\rho_1 = 2v_{1\alpha}^2/(\gamma-1) = 2(\gamma-1)\vphi^2(\sin\alpha_s)^2/(\gamma+1)^2$.  

The post-shock flow is assumed to be self-similar: the flow velocities $v_\varpi,v_\alpha$ and  $c_i$ are functions only of $\alpha$, while the density and pressure are functions of $\alpha$ times $\varpi^n$.     The only radial derivative we require is $\partial \ln \rho/\partial \varpi = n/\varpi$. 

The equations are as follows.  Using the self-similar ansatz, the equation of continuity, ${\mathbf v}\cdot \nabla \ln \rho + \nabla \cdot {\mathbf v}$, becomes 
\begin{equation} \label{continuity} 
(n+1) {v_\varpi\over v_\alpha} + {\partial \over \partial \alpha} \ln (\rho v_\alpha) = 0. 
\end{equation} 
The equation of adiabatic flow, $\vvec \cdot \nabla \ln(c_i^2/\rho^{\gamma-1})=0$, becomes 
\begin{equation} \label{adiabatic} 
{v_\alpha \over c_i^2} \ddalpha c_i^2 - (\gamma-1) n v_\varpi- (\gamma-1) v_\alpha \ddalpha \ln \rho = 0 
\end{equation} 
and this can be combined with equation (\ref{continuity}) to give $\partial {\cal A}/\partial \alpha = 0$ for  
\begin{equation} \label{AngleConservedQty} 
{\cal A} = {c_i^{2(n+1)} v_\alpha^{n(\gamma-1)} \over \rho^{\gamma-1} }. 
\end{equation} 
${\cal A}$ is an angular integral of motion for self-similar, steady, adiabatic flow. 

The acceleration equation is $\vvec \cdot \nabla \vvec  + \nabla c_i^2 + c_i^2 \nabla \ln \rho=0$; its radial component gives 
\begin{equation} \label{acceleration-r} 
v_\alpha \ddalpha v_\varpi -v_\alpha^2 + n c_i^2 = 0 
\end{equation} 
and the azimuthal component gives 
\begin{equation} \label{acceleration-alpha} 
v_\alpha \ddalpha v_\alpha + v_\alpha v_\varpi + \ddalpha c_i^2 + c_i^2 \ddalpha \ln \rho =0. 
\end{equation} 

It is simple to prove that $\vvec \cdot \nabla {\cal B} = 0$, as we expected, in this flow; in our cylindrical coordinates this implies 
\begin{equation} \label{BernoulliCylindrical} 
{\cal B} = \frac12 (v_\varpi^2 + v_\alpha^2) + {\gamma\over \gamma-1} c_i^2 = \frac12 \vphi^2. 
\end{equation} 

 Our flow is defined by the angular dependence of four fluid variables, $v_\varpi,v_\alpha,c_i^2$ and $\ln\rho$, of which two can be eliminated (up to a sign) by the constancy of $\cal A$ and $\cal B$.    In other words, of the four equations 
\begin{equation}\label{vrprime} 
\ddalpha v_\varpi = v_\alpha - n {c_i^2 \over v_\alpha},
\end{equation} 
\begin{equation}\label{valphaprime} 
\ddalpha v_\alpha = -v_\varpi \left[ n{c_i^2\over \gamma c_i^2 - v_\alpha^2} +1\right], 
\end{equation} 
\begin{equation}\label{ci2prime} 
\ddalpha c_i^2 = (\gamma-1) n {v_\varpi \over v_\alpha} {c_i^4  \over \gamma c_i^2 - v_\alpha^2 },
\end{equation} 
and 
\begin{equation}\label{logrhoprime} 
\ddalpha \ln \rho ={n v_\varpi\over v_\alpha}  \left[  { c_i^2 \over \gamma c_i^2 - v_\alpha^2 }-1\right] ,
\end{equation} 
only two are needed.    Taking the ratio of equations (\ref{valphaprime}) and  (\ref{ci2prime}), 
\begin{equation} \label{ci-valpha-ode} 
{ dv_\alpha\over dc_i^2} =  
- {v_\alpha  \over c_i^2} \left[{(\gamma+n)c_i^2 -v_\alpha^2\over n(\gamma-1) c_i^2}\right]
\end{equation} 
and this can be integrated to give $v_\alpha^2(c_i^2)$: 
\begin{equation} \label{valphaSolutionGeneral}
{v_\alpha^2 \over c_i^2} = {n + \gamma(n+2)\over 2 +K \left(c_i^2\right)^{n+\gamma(n+2)\over (\gamma-1)n}} 
\end{equation} 
where $K$ is an integration constant.  Setting $K$ to match the post-shock state, $(\gamma-1)c_{i1}^2 = 2 v_{\alpha1}^2$,  
\begin{equation} \label{valphaSolution}
{v_\alpha^2 \over c_i^2} = {n + \gamma(n+2)\over 2 + 2(n+1){\gamma+1\over \gamma-1} \left(c_i^2/ c_{i1}^2\right)^{n+\gamma(n+2)\over (\gamma-1)n}}; 
\end{equation} 
$v_\varpi$ and $\rho$ can then be derived from ${\cal B}=\vphi^2/2$ and $\partial{\cal A}/\partial\alpha=0$, and the angular variable $\alpha$ can be determined by integrating the inverse of equation (\ref{ci2prime}).   In the physical problem, the flow can only occupy the angular range $\alpha_s<\alpha<\pi$; this places a strong constraint on physically relevant solutions. 

Within solution (\ref{valphaSolution}) the azimuthal Mach number squared, $v_\alpha^2/(\gamma c_i^2)$, ranges from $(\gamma-1)/(2\gamma) <1$ just behind the shock, to $1 + (n/2)(1+1/\gamma) > 1$ for $c_i^2\rightarrow 0$.   The azimuthal sonic point (subscript $s$)  is attained where $\gamma c_{is}^2 = v_{\alpha s}^2$, or 
\begin{eqnarray} \label{ci_SonicPoint_in_Solution}
{c_{is}^2\over c_{i1}^2} &=& \left[ n(\gamma-1)\over 2 \gamma(n+1)\right]^{n(\gamma-1)\over n+\gamma(2+n)}.  \\ 
&~&\nonumber 
\end{eqnarray}

It is clear from equations (\ref{valphaprime})-(\ref{logrhoprime}) that the azimuthal derivatives of $v_\alpha$, $c_i^2$, and $\rho$ diverge at the sonic point unless $v_\varpi=0$ there, so we assume this is true.   Setting $v_\varpi=0$ and $v_\alpha^2 = \gamma c_i^2$ in the Bernoulli equation ${\cal B}=\vphi^2/2$, 
\begin{equation}\label{ci_SonicPoint_Condition} 
c_{is}^2= {\gamma-1 \over \gamma(\gamma+1)} \vphi^2. 
\end{equation}
Combining this with equation (\ref{ci_SonicPoint_in_Solution}) and the post-shock relation $c_{i1}^2/\vphi^2 = 2(\gamma-1)(\sin\alpha_s)^2/(\gamma+1)^2$, we require that the shock angle satisfies 
\begin{equation} \label{ShockAngleSolution} 
(\sin \alpha_s)^2 = {\gamma+1 \over 2\gamma} \left[n(\gamma-1)\over 2\gamma(n+1)\right]^{-{n(\gamma-1)\over n+\gamma(2+n)}}. 
\end{equation} 

Interestingly, this combination of $n$ and $\gamma$ is only less than unity for small values of $n$ (i.e., for $\gamma_p = 1+1/n$ greater than 3.14 when $\gamma=4/3$).   For astrophysically relevant cases, condition (\ref{ShockAngleSolution}) requires $\sin\alpha_s$ to be slightly higher than unity: for instance, when $n=3$, $\gamma=4/3$, it requires $(\sin\alpha_s)^2 = 1.12$.  

This means that one or more of our assumptions must be incorrect.  One possibility is that the small-scale flow is non-steady and also self-similar  in a discrete sense \citep{2009PhFl...21e6101O}.  Another is that the post-shock flow wraps around and affects the upstream fluid, so it is not entirely cold and still in the star's frame prior to the arrival of the shock.  In this case the flow may not in fact be self-similar.    

We draw a few very tentative conclusions.  First, for the shock angle to be as close as possible to the sonic point condition, we expect $\sin \alpha_s \rightarrow 1$, i.e., that the shock meets the surface at a right angle.   Second, we anticipate that the post-shock flow spreads out to to fill the space above the stellar surface, so that streamline angles extend up to $\alpha = \pi/2$.  Third, we may find oscillations in the flow from $|y|<\Lphi$.  All of these issues remain open questions for the numerical investigation of Paper 2. 

\section{Trans-relativistic pattern speeds} \label{S:Transrel}

For a trans-relativistic pattern speed, the oblique flow may still be adiabatic if the explosion is especially vigorous or the star is especially compact.   As in the non-relativistic case, we consider steady flow around the point of breakout.  This flow depends explicitly on the inflow Lorentz factor $\Gammaphi = (1-\vphi^2/c^2)^{-1/2}$, so there is no  universal solution except in the limit $\Gammaphi\gg1$. 

The relativistic Bernoulli function is the ratio of energy flux ($\Gamma^2 w \vvec$) to rest mass flux ($\Gamma \rho_{\rm rest} \vvec$):
\beq\label{Bernoulli-Relativistic}
{\cal B} = \Gamma {w\over \rho_{\rm rest}},
\eeq
where $\Gamma = (1-v^2/c^2)^{-1/2}$ is the flow Lorentz factor, $\rho_{\rm rest}$ is the conserved rest mass density, and $w=p + \rho_{e}$ is the enthalpy, if $\rho_e$ is the total energy density in the fluid frame.  $\cal B$ is conserved along streamlines in steady flow, and this implies a common terminal velocity in the shock frame just as we found in the non-relativistic case.  To see this, note that ${\cal B}=\Gamma^2c^2$ in a region of pressure-free flow; since both the inflow and the terminal outflow are pressure-free, $\Gamma\rightarrow \Gammaphi$ in the outflow (as measured in the steady-state frame). 

The outcome of oblique breakout for significant values of $\Gammaphi$ depends critically on the angle through which streamlines bend to achieve their final directions in the steady-state frame. Suppose the post-shock fluid turns toward a final angle $\alpha_f$ in this frame and accelerates to its terminal Lorentz factor. Here $\alpha_f=\pi$ scorresponds to flowing upstream along the stellar surface, and $\alpha_f=0$ corresponds to no deflection. 

Translated into the star's frame, the ejecta Lorentz factor is $\Gammaphi^2(1-\cos\alpha_f) + \cos \alpha_f$.  Unless $\alpha_f$ is very small ($1-\cos\alpha_f < (\Gammaphi+1)^{-1}$), the ejecta accelerates to Lorentz factors of order $\Gammaphi^2$ (the maximum value being $2\Gammaphi^2-1$, for $\alpha_f=\pi$) and is beamed forward along the stellar surface within an angle of order $1/\Gammaphi$ -- as though it were reflected off a mirror moving with the oblique shock.    

Although this final Lorentz factor is high, it still represents a limitation of the energy budget in very fast ejecta.  In the planar case, shock acceleration continues to Lorentz factors  $\hat \Gamma_s\gg \Gammaphi$ (so long as the flow remains optically thick), and the final Lorentz factor of each fluid element is related to its shock velocity by $\hat \Gamma_f(m) \simeq \hat \Gamma_s(m)^{2.73}$  (\citealt{2001ApJ...551..946T};  see also \citealt{1971PhRvD...3..858J}).  Again, the effect of oblique flow is to end shock acceleration ($\Gamma_s\leq\Gammaphi$) and to deflect the ejecta away from the radial direction.

\begin{deluxetable*}{lccccccl}
\tabletypesize{\scriptsize}
\tablewidth{0pt}
\tablecaption{Model core-collapse supernovae and their minimum eccentricities\tablenotemark{a}  for oblique breakout}
\tablehead{
\colhead{Model} & \colhead{$M_{\rm ej}$ ($M_\odot$) } &
\colhead{$R_*$ ($R_\odot$)} & 
\colhead{$E_{\rm in}$ ($10^{51}$\,erg)} & 
\colhead{$\vsmax$ } & 
\colhead{$R_*/t_{\rm se}$ } & 
\colhead{$\varepsilon_{\rm min}$} & 
\colhead{Reference}
}
\startdata
RSG    & 14 & 500 & 1      & $0.0187 c$ & $0.0117c$ & 0.31	& model s15s7b2 of \citet{1995ApJS..101..181W}\tablenotemark{b} \\
BSG    & 15 & 49   & 1       & $0.075c$ & $0.016c$ & 0.11& \citet{1990ApJ...360..242S}\tablenotemark{b} \\
Ic & 5   & 0.2           & 1      & $0.604c$   & $0.0256c$  & 0.02 & model CO6 of \citet{1999ApJ...516..788W}\tablenotemark{b}\\ 
Ic-BL   & 5      & 0.2  & 30  &$0.984c$ & 0.138c &  0.07& model CO6 of \citet{1999ApJ...516..788W}\tablenotemark{c} 
\enddata
\tablenotetext{a}{Eccentricity: the shock emerges earlier at the pole than the equator by a factor $(1-\varepsilon)/(1+\varepsilon)$. }
\tablenotetext{b}{Provided by Stan Woosley. Model CO6 represents the progenitor of SN 1998bw.}
\tablenotetext{c}{Provided by Ken'ichi Nomoto.  This model represents the progenitor of SN 1987A.} 
\label{Table:Examples}
\end{deluxetable*}

\section{Application to Model Supernovae }\label{S:Examples}

How distorted must an explosion become in order for shock breakout to be strongly affected by obliquity?   Let us address this question in the context of specific models.  Rather than conduct two- or three-dimensional simulations, we consider a few spherical explosions, and apply angle-dependent perturbations to the shock strength ($v_1$) and shock arrival time ($t_{\rm se}$) in each. 

Our four models consist of red and blue supergiant progenitors for Type-II explosions, a compact progenitor for a type Ic supernova, and a high energy broad-lined type Ic explosion from the same progenitor.  All of these progenitor models were kindly provided by Stan Woosley and Ken'ichi Nomoto, and used previously by \citet{1999ApJ...510..379M} and \citet{2001ApJ...551..946T}.  For each we know the stellar mass and radius, as well as the run of density and composition with radius. Choosing an explosion energy and ejected mass for each, we complete our unperturbed models by calculating the run of shock velocity and integrating to find the unperturbed shock emergence time $t_{\rm se,1}$, and by determining the unperturbed shock strength coefficient $v_{1,1}$ and maximum shock velocity $\hat v_{s,{\rm max},1}$.     The model parameters are summarized in Table \ref{Table:Examples}. 

We then consider perturbations to both the shock strength $v_1$ and the shock emergence time $t_{\rm se}$ which are functions of the angle $\theta$ from an arbitrary axis.   To be specific, we 
adopt the patterns which would result from a homologously expanding ellipsoid: 
$ t_{\rm se}(\theta) =(1-\varepsilon \cos2\theta) t_{\rm se,1}$ and $v_1(\theta)= v_{1,1}/(1-\varepsilon\cos2\theta) $.  Here $0\leq \varepsilon<1$ reflects the degree of asphericity; the eccentricity of our reference ellipsoid is $\sqrt{2\varepsilon/(1+\varepsilon)}$.   From these functions we derive the pattern speed $\vphi(\theta) = R_*/[2\varepsilon t_{\rm se,1} \sin 2\theta]$. The angle-dependent shock speed limit $\vsmax(\theta)$ is modified by the factor $(1-\varepsilon\cos2\theta)^{-\delta}$  where $\delta = 1.13$ for the RSG progenitor, and $\delta = 1.17$ for the others. (The difference arises in equation (\ref{vsmax}) from the lower polytropic index $n\simeq 3/2$ in the RSG progenitor.) 

Shock breakout becomes oblique on those portions of the stellar surface where $\vphi<\vsmax$.  
Within each model
\begin{equation}
\frac{\vphi}{\vsmax} = {R_*\over 2\varepsilon v_{s,{\rm max},1} t_{\rm se,1} } {(1-\varepsilon \cos2\theta)^{\delta} \over \sin 2\theta}. 
\end{equation} 
Ignoring the variation in $\vsmax$, which is minor (because $(1-\varepsilon \cos2\theta)^{\delta}\simeq1$), this ratio takes a minimum value of $R_*/(2\varepsilon v_{s,{\rm max},1} t_{\rm se,1})$ at $\theta=\pi/4$ and becomes infinite at the pole and equator of the explosion.  An oblique region therefore exists so long as $\varepsilon \geq \varepsilon_{\rm min}$: 
\begin{equation}
\varepsilon_{\rm min} = {R_*\over 2 v_{s,{\rm max},1} t_{\rm se,1}}. 
\end{equation}
The critical degree of asphericity, $\varepsilon_{\rm min}$, is listed for each model in Table \ref{Table:Examples}.  If $\varepsilon> (2/\sqrt{3}) \varepsilon_{\rm min}$ (an increase of only 15\%), then half or more of the stellar surface is engulfed by oblique flow.   However an explosion must have $\varepsilon \gg \varepsilon_{\rm min}$ for its surface to include the strongly oblique limit $\vphi\ll \vsmax$. 

Two trends are evident in this table.  First, as the progenitor becomes more extended (from Ic to BSG to RSG), a more elongated explosion is required to create oblique flow.  This follows from the fact that $\vsmax$ is approximately proportional to $v_*  R_*^{-0.3}\kappa^{1/6}$, while $R_*/t_{\rm se}$ is approximately proportional to $v_*$, so long as each of these is non-relativistic.  The ratio of these sets $\varepsilon_{\rm min}\propto R^{0.3}\kappa^{-1/6}$  for non-relativistic breakouts.  Because stripped-envelope core-collapse supernovae are intrinsically more asymmetrical than Type II explosions, oblique breakout is much more easily achieved in type Ib and Ic explosions than in Type IIs. 

Second, in the hyper-energetic type Ic explosion (an analogue for SN 1998bw) a higher eccentricity is required to produce oblique breakout than in the lower-energy version of the same progenitor.  The underlying reason is that shock breakout is relativistic in both these explosions, so $\vsmax$ saturates at $c$.    However the stellar explosion is non-relativistic overall, so $R_\star/t_{\rm se}\propto v_*$.  As a result, $\varepsilon_{\rm min}\propto v_*=E_{\rm in}^{1/2} M_{\rm ej}^{-1/2}$ for stars compact enough have relativistic shock breakouts. 

In regions of the stellar atmosphere that are not strongly affected by obliquity ($\vphi \gg \vsmax$), the fastest ejecta are deflected by an amount given by equation (\ref{deflection_angle}), evaluated with $\hat v_s\rightarrow \vsmax$.

\section{Implications for Breakout and External Shock Emission} \label{S:implications} 

Non-radial flows accompany a supernova shock as it traverses the stellar surface, and when the pattern speed is slower than the terminal shock velocity ($\vphi<\vsmax$) these flows alter all the phenomena associated with spherically-symmetric shock breakout.    The alterations  are: 

\noindent {\em - Termination of shock and post-shock acceleration.}  The shock's speed is limited above by $\vphi$, and the maximum ejecta speed is $2\vphi$ (when $\vphi \ll c$).   These upper limits can be well below what a planar shock would produce.  By quenching relativistic flow, an oblique shock breakout can therefore forestall the acceleration of high-energy cosmic rays, the spallation of light nuclei, and the creation of rapid transients which otherwise arise through collisions with circumstellar matter. 

\noindent{\em - Stifling of the breakout flash. }  In a spherical explosion the maximum shock velocity $\vsmax$ and the flash of photons are determined where the shock velocity $\hat v_s(y)$ falls below the effective diffusion speed $c/[3\tau_0(y)]$, where $\tau_0 = \kappa \int_{y_0}^0\rho_0\, dy_0 = \kappa \rho_0 y_0/(n+1)$ is the initial vertical optical depth.  If $\vphi\ll \vsmax$ so that the flow becomes strongly non-radial, obliquity will affect the process. 

How well, in this case, do photons escape the emerging spray of ejecta?  To answer this, we consider the competition of diffusion and advection in the steady-state frame, considering only diffusion along the out-flowing streamlines.   The diffusion speed along each streamline is approximately $c/(3\tau)$ where $\tau(\varpi,\alpha) = \kappa \int_{\varpi}^\infty \rho\, d\varpi$ is the optical depth to infinity.   We evaluate this away from the region of greatest acceleration ($\varpi>\Lphi$), so we can assume each streamline has reached its terminal velocity $\vphi$ and its terminal outflow angle $\alpha$.  Tracing a streamline back to its initial depth $|y_0|$, where matter of density $\rho_0(y_0)$ flows across the shock at speed $\vphi$, mass conservation requires $\rho(\varpi,\alpha) = (d|y_0|/d\alpha)\rho_0/\varpi $.  The $\tau$ integral diverges logarithmically, so we must truncate it at some distance $\varpi_{\rm out}$ set by curvature of the star or the shock conditions, rather than $\infty$.  Eliminating $\rho_0$ in favor of the initial vertical optical depth $\tau_0(y_0)$, we have 
\[ {\tau(\varpi,\alpha) \over \tau_0(y_0)} = (n+1) {d\ln |y_0|\over d\alpha} \ln \left(\varpi_{\rm out}\over \varpi\right). \] 
Using the condition for planar breakout and the scaling $\tau_0\propto y_0^{n+1}\propto \hat v_s^{-\lambda/(n+1)}$, we find $\tau_0(y_0) = [c/(3\vsmax)] (\vsmax/\vphi)^{(n+1)/\lambda} (|y|_0/\Lphi)^{n+1} $.  With this in the above expression, the condition $c/(3\tau)>\vphi$ for photons to diffuse ahead of the flow becomes 
\beq\label{photon_diffusion_criterion} 
{|y_0|\over \Lphi} <
 \left[ d\alpha /d\ln |y_0| \over (n+1) \ln(\varpi_{\rm out}/\varpi)\right]^{1\over n+1}
\left(\vphi\over \vsmax \right)^{{1\over \lambda} - {1\over n+1}}. 
\eeq
For example, if $n=3$, the condition becomes \[ |y_0|/\Lphi < \frac{1}{\sqrt{2}} \left[(d\alpha/d\ln|y_0|)\over \ln(\varpi_{\rm out}/\varpi)\right]^{1/4} \left(\vphi\over\vsmax\right)^{1.54}.\]   
Because the factor $\ln(\varpi_{\rm out}/\varpi)$ is very insensitive to location, and adiabatic expansion saps thermal energy at large distances, we see that when $\vphi\ll \vsmax$, diffusion can only occur on streamlines very close to the stellar surface (e.g., $|y_0|\sim (\vphi/\vsmax)^{1.54}\Lphi$ when $n=3$).
These streamlines curve forward in the post-shock flow, and the considerations in \S~\ref{S:NR-SimilarityAnalysis} imply that they may in fact be trapped along the stellar surface and shielded from all observers by the remainder of the flow.    

This calculation indicates that the direct emission from strongly oblique shock breakouts to be very dim or unobservable, perhaps hardly distinguishable from the phase of decaying luminosity discussed by \citet{1992ApJ...394..599C}, \citet{2010ApJ...725..904N}, and \citet{2011ApJ...728...63R}.\footnote{This phase is sometimes referred to as shock breakout emission, although it  involves spherical expansion and is quite distinct from the breakout flash.} 
This conclusion does not apply to those portions of the stellar surface for which $\vphi \gtrsim \vsmax$, where the shock is essentially normal and obliquity is a small perturbation.  Because the shock strength $v_1$ is likely to correlate with the pattern speed $\vphi$ (for instance, $v_1$ is likely to be greatest where the shock first breaks the surface), this means that the breakout emission is even more strongly localized than we would expect on the basis of spherical theory.  It also means that the breakout flash in an asymmetric explosion can be significantly shorter than the time over which an observer would see the shock cross the stellar surface (the minimum duration identified by \citealt{2004MNRAS.351..694C}, \citealt{2010ApJ...717L.154S}, and \citealt{2011ApJ...727..104C}), as only part of the stellar surface can participate. 

An important caveat to the above calculation is that it considers only diffusion along each streamline.   Photons diffusing across streamlines can potentially make their way from a higher optical depth (we estimate $\tau_0\simeq c/(3\vphi)$) into the thin layer where diffusion is rapid, then out ahead of the ejecta.  There is a possibility that multiple scatterings in the shear region between ejecta and the un-shocked atmosphere would Comptonize these photons.

\noindent{\em - Collisions among non-radial ejecta. }  Regions of oblique breakout can nevertheless produce breakout-related transients through a new channel, because of the non-radial spray of ejecta forward of the advancing shock.  This spray moves above or along the stellar surface at speeds up to twice the pattern speed of the shock (or, in the trans-relativistic case, at Lorentz factors up to $2\Gammaphi^2-1$), and collisions between ejecta sprays should occur wherever oblique shocks advance toward each other.   Because the collision takes place at radii of order a few $R_*$, where the optical depth is decreased, it will be more capable of converting kinetic energy into radiation than was the shock which produced it.    

Comparing the energy content of non-radial ejecta in an oblique breakout ($\vphi \ll \vsmax$) to the energy budget of a planar breakout  ($\vphi \gg \vsmax$) with the same shock strength, we see that the non-radial ejecta receive an energy comparable to what they would have obtained in the planar case.  Part of this energy is available to be re-radiated after ejecta collide. 

\section{Discussion}\label{S:discussion}

The theory presented here raises several questions.  First, have oblique-shock breakouts appeared in prior work?  We believe they have, in those multi-dimensional simulations with sufficient resolution near the stellar surface.   For instance, a lateral spray of ejecta is seen emanating from the emerging jet in the collapsar models of \citeauthor{2008ApJS..176..467W} (2008; their fig.~23).   In the jet-driven supernova simulations of \citet{2011ApJ...727..104C}, a band of dense ejecta forms in the equatorial plane during the phase of free expansion (their figs.~2-6), which we assume is produced by the collision of non-radial ejecta from higher and lower latitudes.  Oblique breakout occurs, albeit strongly modified by gravity, in the propagating neutron star detonation simulations by \citet{2001ApJS..133..195Z}. 

Second, how different are previous predictions of the breakout flash in asymmetrical supernovae from our findings in \S~\ref{S:implications}?  They are quite different.  \citet{2004MNRAS.351..694C} and \citet{2010ApJ...717L.154S} assumed that each patch of the stellar surface would emit as it does in the spherically-symmetric case, but this is only true where $\vphi > \vsmax$.   We find emission to be stifled in regions of the stellar surface for which $\vphi<\vsmax$, although collisions among ejecta provide a second chance for emission.  In \S\ref{S:Examples} we provided concrete examples of the degree of asphericity required for oblique flow to develop within different progenitors, and found type I explosions to be especially susceptible.

\citet{2011ApJ...727..104C} adopt a very different approach, post-processing adiabatic simulations to identify a thermalization photosphere, which they assume emits as a blackbody at the simulation temperature.  This has the benefit that it responds to non-radial flows of ejecta; however, as  \citeauthor{2011ApJ...727..104C}\ acknowledge, it is not 
an especially realistic treatment of radiative transfer.  We expect that similar global calculations, with an improved radiation treatment and careful attention to the resolution of the obliquity scale, will ultimately be very fruitful; however there is much to be gained from simulations focused on the obliquity scale. 

Third, how do our results affect the interpretation of specific events which have early observations or have been associated with shock breakout?  We consider several examples: 

{\em SN 1998bw (GRB 980425) and SN 2003lw (GRB 031203):} These over-energetic type Ic supernovae were both found \citep{1998Natur.395..670G,2004Natur.430..648S} by association with short ($\sim 35$\,s and 20\,s, respectively), smooth-pulse, under-energetic GRBs ($\sim10^{48}$\,erg and $\sim10^{50}$ erg isotropic, respectively).   Off-axis emission from a relativistic jet was proposed to explain GRB 980425 \citep{1999ApJ...521..179H,2003ApJ...594L..79Y}, but neither burst developed the delayed radio afterglow expected in this model \citep{2004ApJ...607L..13S,2004Natur.430..648S}.    Whereas the duration and energy of GRB 980425 are marginally consistent with circumstellar interaction around a spherical explosion of  SN1998bw \citep{1999ApJ...510..379M,2001ApJ...551..946T}, GRB 031203 is a hundred times brighter than a spherical model of SN 2003lw can produce.  Given the strong sensitivity of breakout energetics to asphericity (\S~\ref{S:intro}), it may be possible to explain GRB 031203 as circumstellar interaction following an aspherical, relativistic shock breakout (with some beaming of the emission).   Our results in \S\,\ref{S:Examples} imply that part of the surface of SN 2003lw must, in that scenario, have experienced oblique breakout, so part of its flash could arise from non-radial collisions.   However, the flow is most radial along the axis of a prolate explosion, so the flash in this direction will be least affected by the effects of obliquity. 

{\em SN 2006aj (GRB 060218):} Associated with its smooth-pulse, soft, half-hour-long, $10^{49.8}$-erg $\gamma$-ray burst \citep{2006Natur.442.1008C}, this broad-lined type Ic supernova displayed a high degree of early optical linear polarization in its first few days \citep[4\% for $3<t<5$ d,][]{2007A&A...475L...1M} and in oxygen and iron lines \citep[at 10 d,][]{2006A&A...459L..33G}, which later disappeared in its nebular phase \citep{2006Natur.442.1018M}.  \citet{2007ApJ...667..351W} interpret GRB 060218 as the emergence of a semi-relativistic shock from the photosphere of a thick wind.  High early polarization, and the fact that the burst energy is too great to explain within a spherical model, imply a strongly asymmetric breakout in the compact progenitor.  We note that any transient produced by non-radial ejecta would have occurred on a time scale of order a few $R_*/\vphi$, briefer than the observed burst.  If such a collision did occur, it was likely hidden beneath the opaque wind. 

{\em SN 2008D (XRF 080109):} Identified through its $10^{46.3}$ erg, several-hundred-second X-ray flash, this type Ib supernova showed trans-relativistic initial expansion \citep{2008Natur.453..469S}.  The $\sim 1\%$ intrinsic polarization \citep{2009ApJ...705.1139M,2011ASPC..449..421G} and spectral lines from the deep oxygen ejecta \citep{2009ApJ...702..226M} indicate some degree of asymmetry, and \citet{2011ApJ...727..104C} invoke asymmetrical breakout to explain the flash duration.   No direct breakout flash is expected where the shock's motion across the surface is slow enough to make the breakout strongly oblique, a point which makes this explanation less likely.  The flash could instead have arisen indirectly, from the collision of non-radial ejecta. 

{\em SN 2010jp:} \citet{2012MNRAS.420.1135S} interpret triple-peaked H$\alpha$ lines in this dim, peculiar type IIn supernova  as evidence of jet-like lobes in the high-velocity ejecta, which along with the very low mass of ejected $^{56}$Ni suggest a jet-driven explosion.    Simulations of jet-driven explosions \citep[e.g.,][]{1999ApJ...524..262M,2000ApJ...537..810W,2001AIPC..586..459H,2011ApJ...727..104C} show that jets are typically stopped and contained within extended, hydrogen-rich envelopes like that of SN 2010jp's progenitor; see \citet{2003MNRAS.345..575M} for analytical criteria on this point.  Strong asymmetry in the high-velocity ejecta suggest its breakout may have become oblique, triggering non-radial flows and possibly an equatorial band of shocked, non-radial ejecta.  However, as noted above, oblique flow is more difficult to achieve in an extended star, like the presumed progenitor of SN 2010jp, than in a compact one.  In particular, the models outlined in Table \ref{Table:Examples} would require ellipticities significantly above 11\% or 31\% for blue and red supergiant progenitors, respectively, corresponding to equatorial breakouts which lag the pole by time factors of 1.24 and 1.90, respectively.

{\em SN 2011dh:} This type IIb explosion in M51 is now known to have originated from a yellow supergiant (YSG) progenitor which has since disappeared \citep{2013arXiv1305.3436V}, whereas it was initially identified with a much more compact star 
\citep{2011ApJ...742L..18A,2012ApJ...752...78S} on the basis of its early luminosity and radio emission.   The discrepancy is a challenge for spherical theory, which robustly predicts the early luminosity evolution of supernovae \citep{1992ApJ...394..599C,2011ApJ...728...63R}.   Aspherical motions of the high-velocity ejecta may resolve it, if it is possible for the observed expansion speed ($[2.1\pm0.7]\times 10^4$\,km\,s$^{-1}$ according to \citealt{2012ApJ...751..125B}, or $[1.5\pm0.18]\times 10^4$\,km\,s$^{-1}$ according to \citealt{2012arXiv1209.1102H}) to reflect the oblique upper limit  $2\vphi$ rather than the planar upper limit $C_2 \vsmax$ set by radiation diffusion.  The spherical upper limit is approximately $\vsmax = 2.0\times10^4\,(E_{\rm in}/10^{51}\,{\rm erg})^{0.58}$\,km\,s$^{-1}$ in \citet{1999ApJ...510..379M}'s spherical model for this YSG explosion.

\section{Conclusions}\label{Conclusions} 

Avenues for future work include numerical simulations to complete our picture of the strongly oblique limit for non-relativistic pattern speeds (Paper 2), and relativistic and superluminal pattern speeds as well.  With radiation hydrodynamical codes it will be possible to survey the critical parameter $\vphi/\vsmax$ and the transition from weak to strong obliquity.  The possibility that non-radial flows lead to collisions should be addressed with high-resolution, multi-dimensional simulations. 

We conclude with a couple general comments on the consequences of a change in the dynamics and emission from shock emergence in supernovae. 

Obliquity not only alters the emission at the time of shock emergence, but also changes the distribution of matter and heat across velocity (and angle) in the highest-velocity ejecta. This is an important consideration when the early luminosity evolution \citep[e.g.,][]{1992ApJ...394..599C}, early radio emission \citep[e.g.,][]{2013ApJ...762...14M}, or direct breakout emission \citep[e.g.,][]{2004MNRAS.351..694C} are used to constrain the stellar radius, shock dynamics, or circumstellar environment, or when collisions with circumstellar disks \citep{2010MNRAS.409..284M}, companion stars \citep{2010ApJ...708.1025K}, or nearby proto-planetary systems \citep{2007ApJ...662.1268O} are considered. 

Finally, oblique shock breakouts are not restricted to core-collapse supernovae and jet-driven $\gamma$-ray bursts.  They are also expected, and have potentially observable consequences, in Type Ia supernova explosions of white dwarfs \citep{2010ApJ...708..598P}, accretion-induced collapses of white dwarfs \citep{1999ApJ...516..892F}, subsurface detonations in white dwarfs and neutron stars \citep{2007ApJ...670.1291W,2012ApJ...755....4T}, and stellar tidal disruptions by massive black holes \citep{2004ApJ...615..855K,2009ApJ...705..844G}.  

%%%%%%%%%%%%%%%%%%%%%%%%%%%%%%%%%%%%%%%%%%%%%%%%%%%%%%%%%%%%%%%%%%%%%%%%%%%%%%%%%%%%%%%%%%%%%%%%%%%%%%%%%

\section*{}
CDM and SR are supported by NSERC through a Discovery Grant.  YL's research is supported by an ARC Future Fellowship.   The authors are indebted to the referee for useful clarifications, to Eli Waxman for a discussion of shock breakout emission, and to Marten van Kerkwijk for bringing SN 2011dh to our attention.  CDM thanks Ken'ichi Nomoto and Stan Woosley for generously providing progenitor models. 

%%%%%%%%%%%%%%%%%%%%Begin the Reference%%%%%%%%%%%%%%%%%%%%%%%%%%

%%%%%%%%%%%%%%%%%%%%End the Reference%%%%%%%%%%%%%%%%%%%%%%%%%%%

%%%%%%%%%%%%%%%%%%%%%%%%%%%%%%%Begin the Appendix%%%%%%%%%%%%%%%%%%%%%%%%%%%%%%%%%%%%%%%%%%%%%%%%%%
%\appendix
%
%
%%%%%%%%%%%%%%%%%%%%%%%%%%%%%%%%Begin Appendix B%%%%%%%%%%%%%%%%%%%%%%%%%%%%%
%\section{A. }\label{Sec:AppendixA}

\end{document}